\documentclass[aps,onecolumn,showpacs,nofootinbib,showkeys]{revtex4-2}
\usepackage[margin=2.5 cm]{geometry}

\RequirePackage[T1]{fontenc}
 
\usepackage{epsfig,graphicx,amsmath,amssymb,bm}
\RequirePackage{mathptmx}      
\usepackage{mathrsfs}
\usepackage{amssymb}
\RequirePackage{color}
\usepackage{changes}
\usepackage{slashed}
\usepackage{multirow}
\usepackage{scalerel}
\usepackage{tikz-feynman}
\usepackage{textcomp}
\usepackage{subcaption}
\usepackage[english]{babel}
\usepackage{float}
\RequirePackage{hyperref}
\hypersetup{
    linktocpage,
    colorlinks,
    citecolor=blue,
    filecolor=black,
    linkcolor=blue,
    urlcolor=blue,
}
\usepackage{nccmath}

\begin{document}

\title{
$\tau$ lepton decays with production of strange scalar mesons $K^*_0(700)$ and $K^*_0(1430)$ in the extended NJL model
}


\author{Mikhail K. Volkov$^{1}$}\email{volkov@theor.jinr.ru}
\author{Kanat Nurlan$^{1,2,3}$}\email{nurlan@theor.jinr.ru}

\affiliation{$^1$ Bogoliubov Laboratory of Theoretical Physics, JINR, 
                 141980 Dubna, Moscow region, Russia \\
                $^2$ The Institute of Nuclear Physics, Almaty, 050032, Kazakhstan\\
                $^3$ L. N. Gumilyov Eurasian National University, Astana, 010008, Kazakhstan}   


\begin{abstract}
The branching fractions of $\tau$ lepton decays with the production of strange scalar mesons both in the ground and first radially excited states $\tau \to \nu_\tau [K^*_0(700), K^*_0(1430)]$ and $\tau \to \nu_\tau [K^*_0(700)\pi, K^*_0(1430)\pi, K^*_0(700) K, K^*_0 (700)\eta]$ are calculated in the extended $U(3) \times U(3)$ chiral quark NJL model. All mesons are considered as $q\bar{q}$ systems. The obtained results should be considered as predictions for future experiments.


\end{abstract}


\maketitle


\section{\label{Intro}Introduction}
The versions of the $U(3) \times U(3)$ chirally symmetric Nambu--Jona-Lasinio (NJL) model are a good tool for the study of light mesons interactions at low-energies \cite{Nambu:1961tp, Eguchi:1976iz, Ebert:1982pk, Volkov:1984kq, Ebert:1985kz, Volkov:1986zb, Vogl:1991qt, Klevansky:1992qe, Hatsuda:1994pi, Ebert:1994mf, Volkov:1999yi, Buballa:2003qv, Volkov:2005kw, Volkov:2017arr, Volkov:2022jfr}. The meson nonets of pseudoscalar, vector, and axial-vector types are considered as quark-antiquark structures in these models. However, when considering the scalar meson nonet, difficulties are encountered in describing the mass of the isovector meson $a_0(980)$ \cite{Volkov:2005kw, Volkov:1998ax}. In addition, problems arise when describing the decays widths of $f_0(980) \to \pi\pi$ and $a_0(980) \to \pi \eta$, where the first width within the $q\bar{q}$ structure has an underestimated value, and the second width under the same assumptions exceeds the experimental value \cite{Volkov:1998ax, tHooft:2008rus}. At present, these difficulties are overcome by introducing additional components into the $f_0(980)$ and $a_0(980)$ mesons in the form of a kaon molecule or in the form of an additional tetraquark structure as leading components compared to $q\bar{ q}$ \cite{tHooft:2008rus, Weinstein:1982gc, Lee:2013mfa, Achasov:1999wv, Ahmed:2020kmp}. A number of other scalar mesons are quite satisfactorily described as a quark-antiquark system. This primarily applies to the $f_0(500)$ meson, strange mesons $K^*_0(700)$ and $K^*_0(1430)$, as well as mesons in the first radially excited states \cite{Volkov:2005kw, Volkov:1998ax}. It should also be noted that the existence of a $q\bar{q}$ system for scalar mesons, together with a similar structure for pseudoscalar mesons, plays an important role in ensuring approximate chiral symmetry of strong interactions.

In our recent paper \cite{Volkov:2022ukj}, it was shown that the decays of strange scalar mesons are quite satisfactorily described as a $q\bar{q}$ system in agreement with recent experiments by the BaBar collaboration \cite{BaBar:2021fkz}. This concerns the main decays of the strange scalar meson in the ground $K^*_0(700) \to K \pi$ and first radially excited states $K^*_0(1430) \to K\pi, K\eta, K \eta', K_1 \pi$. In the present paper, we continue these studies and give predictions for the decays of $\tau$ leptons with the production of strange scalar mesons: $\tau \to \nu_\tau [K^*_0(700), K^*_0(1430)]$ and $\tau \to \nu_\tau [K^*_0( 700)\pi, K^*_0(1430)\pi, K^*_0(700) K, K^*_0(700)\eta]$.
 
\section{Effective quark-meson Lagrangians of the NJL model}
A distinctive feature of our $U(3) \times U(3)$ version of the NJL chiral model is a relatively large value of the cutoff parameter equal to $\Lambda_4 =1260$ MeV \cite{Volkov:1986zb, Volkov:2005kw}. This value is commensurable in its magnitude not only with the masses of the 4 meson nonets (scalar, pseudoscalar, vector and axial-vector) but also with the masses of their first radially excited states. This gives grounds for attempts to include mesons in the first radially excited states into the model without violating the partial conservation of chiral symmetry. The model of this type was constructed in \cite{Volkov:1996br, Volkov:1996fk}. At the same time, many numerical calculations of various processes involving the first radially excited mesons have shown satisfactory agreement between the obtained results in the model and experimental data \cite{Volkov:1999yi, Volkov:2017arr, Volkov:2022jfr}. This allows us to hope for the reliability of the obtained results in this work.

In our extended version of the NJL model, the first radially excited mesons are described by introducing the simplest form factor in the polynomial form of the second degree in the relative momentum of quarks. The numerical coefficient at $k^2$ (slope parameter) is fixed from the requirement to preserve the main parameters of the initial standard model. Such parameters are the quark masses and the cutoff parameter \cite{Volkov:1996br, Volkov:1996fk}. 

The next step in constructing the model is the requirement to diagonalize the free Lagrangian, since after the introduction of excited states, nondiagonal terms associated with possible transitions between the ground and excited states appear in this model. Diagonalization is achieved by introducing mixing angles. As a result, when describing the interaction in the new model, there are no more arbitrary parameters than in the initial standard model \cite{Volkov:1986zb, Volkov:1996fk, Volkov:2005kw}.

The complete Lagrangians for the interaction of mesons with quarks are given in \cite{Volkov:1999yi, Volkov:2005kw, Volkov:2017arr, Volkov:2022jfr}. Part of the Lagrangian of strong interactions of scalar, pseudoscalar, vector and axial-vector mesons with quarks necessary for describing processes considered here takes the form
	\begin{eqnarray}
	\label{Lagrangian}
		{\cal L}_{int} & = &
		\bar{q} \biggl[ i \gamma_5 \sum_{i = \pm, 0} \lambda^\pi_i \left( a_\pi \pi^i + b_\pi \hat{\pi}^i \right) + i \gamma_5  \sum_{i = \pm, 0} \lambda^K_i \left( a_K K^i +b_K \hat{K}^i \right) \nonumber \\ 
		&& + \frac{1}{2} \gamma_\mu \gamma_5 \sum_{i = \pm, 0} \lambda^K_i \left( a_{K_1} K^i_{1\mu} + b_{K_1} \hat{K}^i_{1\mu} \right) 
+ \frac{1}{2} \gamma_\mu \gamma_5 \sum_{i = \pm, 0} \lambda^{a_1}_i \left( a_{a_1} a^i_{1\mu} + b_{a_1} \hat{a}^i_{1\mu} \right) \nonumber \\ 
		&&  + \sum_{i = \pm} \lambda_{K}(a_{K^*_0} K^{*i}_0 + b_{K^*_0} \hat{K}^{*i}_0) + i\gamma^{5} \sum_{i = u, s} \lambda_{i} \left[A^{i}_{\eta}\eta + A^{i}_{\eta'}\eta' + A^{i}_{\hat{\eta}}\hat{\eta} + A^{i}_{\hat{\eta}'}\hat{\eta}'\right]
		\biggl]q,
	\end{eqnarray}
where $q$ and $\bar{q}$ are the $U(3)$ triplets of the $u$, $d$ and $s$ quark fields with constituent quark masses $m_{u} \approx m_{d} = 270$~MeV, $m_{s} = 420$~MeV; excited mesonic states of mesons are marked with a hat and $\lambda$ are linear combinations of the Gell-Mann matrices \cite{Volkov:2022jfr},
\begin{eqnarray}
\label{verteces1}
	a_{M} = a^0_{M} \left[g_{M}\sin(\theta_M + \theta^0_M) +
	g'_{M}f_{M}(k_{\perp}^{2})\sin(\theta_M - \theta^0_M) \right], \nonumber\\
	b_{M} = - a^0_{M} \left[g_{M}\cos(\theta_M + \theta^0_M) +
	g'_{M}f_{M}(k_{\perp}^{2})\cos(\theta_M + \theta^0_M) \right],
\end{eqnarray}
where $a^0_{M} = 1/{\sin(2\theta_{M}^{0})}$. The subscript M indicates the corresponding meson. The mixing angles are given in Table \ref{tab_mixing}. The mixing angles for the $K$ and $\pi$ mesons $\theta \approx \theta_0$; so for the ground states of these mesons one can use $A_\pi = g_\pi$ and $A_K=g_K$.
\begin{table}[h!]
\begin{center}
\begin{tabular}{cccccc}
\hline
   & $\pi$ & $K$ & $a_1$ & $K_1$ & $K^*_0$ \\
\hline
$\theta_M$	& $59.48^{\circ}$	& $58.11^{\circ}$     & $81.80^{\circ}$    & $85.97^{\circ}$   & $74.0^{\circ}$  \\
$\theta^0_M$	& $59.12^{\circ}$	& $55.52^{\circ}$     & $61.50^{\circ}$    & $59.56^{\circ}$   & $60.0^{\circ}$ \\
\hline
\end{tabular}
\end{center}
\caption{Values of the mixing angles for mesons in the ground and first radially excited states \cite{Volkov:1999yi, Volkov:2022jfr}.}
\label{tab_mixing}
\end{table}  

For the $\eta$ mesons, the factor $A$ takes a slightly different form. This is due to the fact that 
in the case four states $\eta, \eta', \eta(1295)$ and $\eta(1475)$ are mixed 
        \begin{eqnarray}
            A^{u}_{M} & = & g_{\eta^{u}} a^{u}_{1M} + g'_{\eta^{u}} a^{u}_{2M} f_{uu}(k_{\perp}^{2}), \nonumber\\
            A^{s}_{M} & = & g_{\eta^{s}} a^{s}_{1M} + g'_{\eta^{s}} a^{s}_{2M} f_{ss}(k_{\perp}^{2}).
        \end{eqnarray}
        
Here $f\left(k_{\perp}^{2}\right) = \left(1 + d k_{\perp}^{2}\right)\Theta(\Lambda^{2} - k_{\perp}^2)$ is 
the form-factor describing the first radially excited meson states. The slope parameters, 
$d_{uu} = -1.784 \times 10^{-6} \textrm{MeV}^{-2}$, $d_{us} = -1.761 \times 10^{-6} \textrm{MeV}^{-2}$ and $d_{ss} = -1.737 \times 10^{-6} \textrm{MeV}^{-2}$, 
are unambiguously fixed from the condition of constancy of the quark condensate after the inclusion 
of radially excited states and depend only on the quark composition of the corresponding meson \cite{Volkov:2022jfr}.

The values of the mixing ($A$) parameters are shown in Table \ref{tab_eta}. The $\eta'$ meson corresponds to the physical state $\eta'(958)$ and the $\hat{\eta}$, $\hat{\eta}'$ mesons correspond to the first radial excitation of the mesons $\eta$ and $\eta'$.
    
\begin{table}[h!]
\begin{center}
\begin{tabular}{ccccc}
\hline
   & $\eta$ & $\hat{\eta}$ & $\eta'$ & $\hat{\eta}'$ \\
\hline
$a^{u}_{1}$		& 0.71			& 0.62            &-0.32             & 0.56    \\
$a^{u}_{2}$		& 0.11			& -0.87           & -0.48            & -0.54   \\
$a^{s}_{1}$               & 0.62                        & 0.19            & 0.56             & -0.67 \\
$a^{s}_{2}$               & 0.06                       & -0.66           & 0.3               & 0.82 \\
\hline
\end{tabular}
\end{center}
\caption{Mixing parameters of $\eta$ mesons \cite{Volkov:1999yi, Volkov:2022jfr}.}
\label{tab_eta}
\end{table}   
    
The quark-meson coupling constants have the form
\begin{eqnarray}
	\label{Couplings}
 g_{\pi} = g_{\eta^{u}}=\left(\frac{4}{Z_{\pi}}I_{20}\right)^{-1/2}, 
\, g'_{\pi}=g'_{\eta^{u}} =  \left(4 I_{20}^{f^{2}}\right)^{-1/2}, 
\, g_{\eta^{s}}=\left(\frac{4}{Z_{\eta^s}}I_{02}\right)^{-1/2}, 
\, g'_{\eta^{s}} =  \left(4 I_{02}^{f^{2}}\right)^{-1/2}, \nonumber\\
\, g_{K} =\left(\frac{4}{Z_K}I_{11}\right)^{-1/2},
\, g'_{K} =\left(4I^{f^2}_{11}\right)^{-1/2},  
 g_{K^*_0} =\left(4I_{11}\right)^{-1/2}, 
\, g'_{K^*_0} =\left(4I^{f^2}_{11}\right)^{-1/2}, \nonumber\\
g_{a_1} =\left(\frac{2}{3}I_{20}\right)^{-1/2}, 
\, g'_{a_1} =\left(\frac{2}{3}I_{20}^{f^{2}}\right)^{-1/2}, 
g_{K_1} =\left(\frac{2}{3}I_{11}\right)^{-1/2}, 
\, g'_{K_1} =\left(\frac{2}{3}I_{11}^{f^{2}}\right)^{-1/2}, 
\end{eqnarray}
here $Z_{\pi}$ and $Z_{\eta^{s}}$ are additional renormalization constants appearing 
in the transitions between pseudoscalar and axial-vector mesons \cite{Volkov:2005kw, Volkov:2022jfr}.

Integrals appearing in the quark loops are
\begin{eqnarray}
	I_{n_{1}n_{2}}^{f^{m}} =
	-i\frac{N_{c}}{(2\pi)^{4}}\int\frac{f^{m}(k^2_{\perp})}{(m_{u}^{2} - k^2)^{n_{1}}(m_{s}^{2} - k^2)^{n_{2}}}\Theta(\Lambda_{3}^{2} - k^2_{\perp})
	\mathrm{d}^{4}k.
\end{eqnarray}
where $ \Lambda_3=1030$ MeV is the three-dimensional cutoff parameter, the value of the four-dimensional cutoff parameter is $\Lambda_4=1260$ MeV \cite{Volkov:2005kw}. 

When describing decays involving $K_1$ axial vector mesons, we take into account the mixing effect of the $K_{1A}$ and $K_{1B}$ states \cite{Volkov:1986zb, Suzuki:1993yc}. The mixing of the axial vector mesons $K_{1A}$ and $K_{1B}$ leads to physical states $K_1 (1270)$ and $K_1 (1400)$ \cite{ParticleDataGroup:2022pth}. This mixing is described as follows:
	\begin{eqnarray}
	\label{K1AK1B}
		K_{1A} & = & K_{1}(1270)\sin{\alpha} + K_{1}(1400)\cos{\alpha}, \nonumber\\
		K_{1B} & = & K_{1}(1270)\cos{\alpha} - K_{1}(1400)\sin{\alpha},
	\end{eqnarray}
where $\alpha=57^{\circ}$	\cite{Volkov:2022jfr}. This effect was also considered in the works \cite{Volkov:1984gqw, Li:1996md, Li:2005eq, Geng:2006yb, Cheng:2013cwa}.  

\section{Amplitudes and widths of $\tau$ lepton decays} 
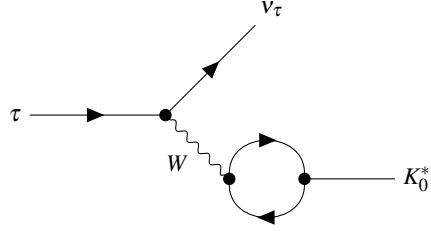
\begin{figure*}[t]
 \centering
  \centering
   \begin{tikzpicture}
    \begin{feynman}
      \vertex (a) {\(\tau\)};
      \vertex [dot, right=2cm of a] (b){};
      \vertex [above right=2cm of b] (c) {\(\nu_{\tau}\)};
      \vertex [dot, below right=1.2cm of b] (d) {};
      \vertex [dot, right=1.0cm of d] (l) {};
      \vertex [right=1.5cm of l] (g) {\(K^*_0\)};      
      \diagram* {
         (a) -- [fermion] (b),
         (b) -- [fermion] (c),
         (b) -- [boson, edge label'=\(W\)] (d),
         (d) -- [fermion, inner sep=1pt, half left] (l),
         (l) -- [fermion, inner sep=1pt, half left] (d),         
         (l) -- [] (g),
      };
     \end{feynman}
    \end{tikzpicture}
 \caption{Diagrams describing the decay $\tau \to K^*_0 \nu_\tau$.}
 \label{diagram1}
\end{figure*}

We start with considering the simplest $\tau$ lepton decay into the $K^*_0(700)$ and $K^*_0(1430)$ mesons. These decays occur due to the difference between the masses of the $u$ and $s$ quarks. In the NJL model, these processes are described by the quark diagram shown in Figure \ref{diagram1}. When calculating this diagram, we expand the integral over the quark loop in terms of the momentum of external fields and retain only the divergent parts \cite{Volkov:1986zb, Volkov:2017arr, Volkov:2022jfr}. As a result, we obtain the following formula for the decay width:
	\begin{eqnarray}
	\label{width_1}
        \Gamma(\tau \to K^*_0(700) \nu_{\tau})_{NJL}=
        \frac{\sqrt{E^2_{K^*_0} - M^2_{K^*_0}}}{4 \pi} \left(M^2_{\tau} - M^2_{K^*_0} \right) {\biggl[ G_F V_{us} \frac{m_s - m_u}{2} \frac{C_{K^*_0}}{g_{K^*_0}} \biggl]}^2,
	\end{eqnarray}
where the $\tau$ lepton and meson masses are taken from PDG \cite{ParticleDataGroup:2022pth}; $E_{K^*_0} = M^2_{\tau}+M^2_{K^*_0} / 2M_{\tau}$ is the energy of the scalar meson $K^*_0$ in the rest frame of the $\tau$ lepton; $G_{F}$ is the Fermi constant and $V_{us}$ is the element of the Cabibbo-Kobayashi-Maskawa matrix; $C_{K^*_0} = 0.97$ is the factor arising from the quark loops describing the $W$ boson transition to the $K^*_0$ meson \cite{Volkov:2022jfr}. In the NJL model, we obtain $Br(\tau \to K^*_0(700) \nu_\tau)_{NJL} = 4.87 \times 10^{-4}$ for the branching fraction of this decay.  

The decay width of $\tau \to K^*_0(1430) \nu_\tau$ can be obtained from formula (\ref{width_1}) by replacing the mass $M_{K^*_0} \to M_{K^*_0(1430 )} $ and constants $C_{K^*_0} \to C_{K^{*'}_0}$, where $C_{K^{*'}_0} = 0.25$ \cite{Volkov:2022jfr}. As a result, for the branching fraction we obtain $Br(\tau \to K^*_0(1430) \nu_\tau)_{NJL} = 4.07 \times10^{-6}$. The experiment gives only the upper limit for the partial decay width $Br(\tau \to K^*_0(1430) \nu_\tau) < 5 \times10^{-4}$ \cite{ParticleDataGroup:2022pth}. Our result does not exceed this limit. Moreover, we can compare our results with the theoretical results obtained within the model with parameterization of scalar $\pi K$ from factor  $Br(\tau \to K^*_0(1430) \nu_\tau) = (0.07(3) - 0.78(38))\times10^{-4}$ \cite{VonDetten:2021rax}. Our results also do not contradict these theoretical data.

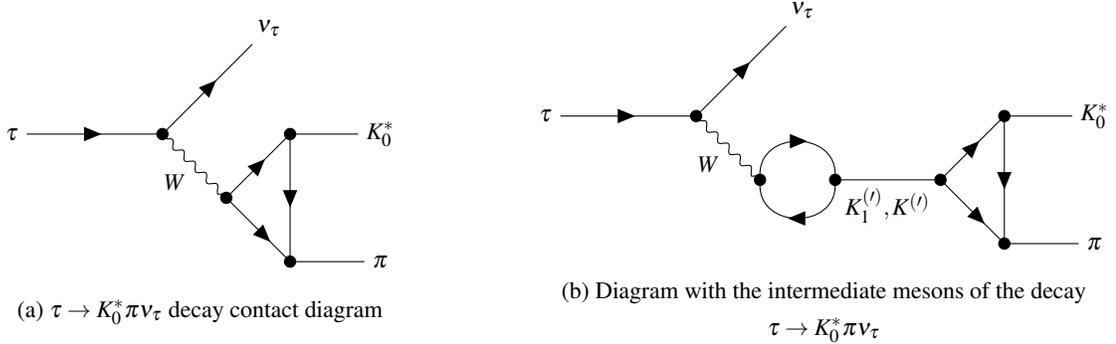
\begin{figure*}[t]
 \centering
  \begin{subfigure}{0.5\textwidth}
   \centering
    \begin{tikzpicture}
     \begin{feynman}
      \vertex (a) {\(\tau\)};
      \vertex [dot, right=2cm of a] (b){};
      \vertex [above right=2cm of b] (c) {\(\nu_{\tau}\)};
      \vertex [dot, below right=1.2cm of b] (d) {};
      \vertex [dot, above right=1.2cm of d] (e) {};
      \vertex [dot, below right=1.2cm of d] (h) {};
      \vertex [right=1.2cm of e] (f) {\(K^*_0\)};
      \vertex [right=1.2cm of h] (k) {\(\pi\)}; 
      \diagram* {
         (a) -- [fermion] (b),
         (b) -- [fermion] (c),
         (b) -- [boson, edge label'=\(W\)] (d),
         (d) -- [fermion] (e),  
         (e) -- [fermion] (h),
         (h) -- [anti fermion] (d),
         (e) -- [] (f),
         (h) -- [] (k),
      };
     \end{feynman}
    \end{tikzpicture}
   \caption{$\tau \to K^*_0 \pi \nu_\tau$ decay contact diagram}
  \end{subfigure}%
 \centering
 \begin{subfigure}{0.5\textwidth}
  \centering
   \begin{tikzpicture}
    \begin{feynman}
      \vertex (a) {\(\tau\)};
      \vertex [dot, right=2cm of a] (b){};
      \vertex [above right=2cm of b] (c) {\(\nu_{\tau}\)};
      \vertex [dot, below right=1.2cm of b] (d) {};
      \vertex [dot, right=1cm of d] (l) {};
      \vertex [dot, right=1.4cm of l] (g) {};
      \vertex [dot, above right=1.2cm of g] (e) {};
      \vertex [dot, below right=1.2cm of g] (h) {};
      \vertex [right=1.2cm of e] (f) {\(K^*_0\)};
      \vertex [right=1.2cm of h] (k) {\(\pi\)}; 
      \diagram* {
         (a) -- [fermion] (b),
         (b) -- [fermion] (c),
         (b) -- [boson, edge label'=\(W\)] (d),
         (d) -- [fermion, inner sep=1pt, half left] (l),
         (l) -- [fermion, inner sep=1pt, half left] (d),
         (l) -- [edge label'=\({ K^{(\prime)}_1, K^{(\prime)}} \)] (g),
         (g) -- [fermion] (e),  
         (e) -- [fermion] (h),
         (h) -- [anti fermion] (g),
         (e) -- [] (f),
         (h) -- [] (k),
      };
     \end{feynman}
    \end{tikzpicture}
   \caption{Diagram with the intermediate mesons of the decay $\tau \to K^*_0 \pi \nu_\tau$}
  \end{subfigure}%
 \caption{Diagrams contributing to the decay $\tau \to K^*_0 \pi \nu_\tau$.}
 \label{diagram2}
\end{figure*}%

Next we consider the $\tau$ lepton decays with the production of $K^*_0(700) \pi$ and $K^*_0(1430) \pi$ meson pairs. For these decays, we take into account the contact channel, when mesons are directly produced through the $W$ boson and the channels with axial vector and pseudoscalar mesons in the ground and first radially excited states. The corresponding diagrams are shown in Figure \ref{diagram2}. As a result, for the total decay amplitude $\tau \to K^*_0(700) \pi \nu_\tau$ we obtain
	\begin{eqnarray}
	\label{amplitude_2}
         \mathcal{M}(\tau \to K^*_0(700) \pi \nu_\tau) = 2G_F V_{us} L_\mu {\biggl[ \mathcal{M}_C + \mathcal{M}_{K_1 + K'_1} + \mathcal{M}_{K + K'} \biggl]}_{\mu\nu} {\left( p_{K^*_0} - p_\pi \right)}_\nu, 
	\end{eqnarray}
where $L_{\mu}$ is the lepton current, $p_{K^*_0}$ and $p_\pi$ are the momenta of the scalar meson $K^*_0(700)$ and pion. 
The contribution from the contact diagram is described by the logarithmic divergent integral ${\mathcal{M}_C}_{\mu\nu} = I^{K^*_0 \pi}_{11} g_{\mu\nu}$. Diagrams with intermediate axial vector mesons $K_1(1270, 1400)$ and $K_1(1650)$ and pseudoscalar mesons $K$ and $K(1460)$ give the following contribution to the total amplitude:  
	\begin{eqnarray}
	\label{amplitude_3}
{\mathcal{M}_{K_1 + K'_1}}_{\mu\nu} & = & \frac{C_{K_1}}{g_{K_1}} I^{K_1 K^*_0 \pi}_{11} \left[ \biggl[ g_{\mu\nu} \left( p^2 - \frac{3}{2} (m_s+m_u)^2 \right) - p_\mu p_\nu \left( 1- \frac{3}{2} \frac{(m_s+m_u)^2}{M^2_{K_1(1270)}} \right) \biggl] BW_{K_1(1270)} \sin^2(\alpha)  \right. \nonumber\\
        && \left. + \biggl[ g_{\mu\nu} \left( p^2 - \frac{3}{2} (m_s+m_u)^2 \right) - p_\mu p_\nu \left( 1- \frac{3}{2} \frac{(m_s+m_u)^2}{M^2_{K_1(1400)}} \right) \biggl] BW_{K_1(1400)} \cos^2(\alpha) 
\right]  \nonumber\\
        && + \frac{C_{K'_1}}{g_{K_1}} I^{K'_1 K^*_0 \pi}_{11} \biggl[ g_{\mu\nu} \left( p^2 - \frac{3}{2} (m_s+m_u)^2 \right) - p_\mu p_\nu \left( 1- \frac{3}{2} \frac{(m_s+m_u)^2}{M^2_{K_1(1650)}} \right) \biggl] BW_{K_1(1650)},
	\end{eqnarray}
	\begin{eqnarray}
	\label{amplitude_4}
{\mathcal{M}_{K + K'}}_{\mu\nu} & = & 4 m_s F_K \biggl[I^{K K^*_0 \pi}_{11} BW_K  + \gamma_{K'} I^{K' K^*_0 \pi}_{11} BW_{K'} \biggl]g_{\mu\nu} p_\nu,
	\end{eqnarray}
where $\gamma_{K'}= F_{K'}/F_K \approx 0.26$ \cite{Volkov:2022jfr}; $C_{K_1} =0.90$ and $C_{K'_1}=0.42$ \cite{Volkov:2022jfr}; $p = p_{K^*_0} + p_\pi$. The integrals
\begin{eqnarray}
\label{integral}
&& I_{n_1n_2}^{M M'...}(m_{u}, m_{s}) = -i\frac{N_{c}}{(2\pi)^{4}} 
 \int\frac{a(k_{\perp}^{2})...b(k_{\perp}^{2})...}{(m_{u}^{2} - k^2)^{n_1}(m_{s}^{2} - k^2)^{n_2}}
\Theta(\Lambda_{3}^{2} - \vec{k}^2) \mathrm{d}^{4}k,
\end{eqnarray}
are obtained from the quark triangular loops, $a(k_{\perp}^{2})$ and $b(k_{\perp}^{2})$ are the coefficients
for different mesons defined in (\ref{verteces1}). Intermediate mesons are described by Breit-Wigner propogators
\begin{eqnarray}
\label{integral}
&& BW_{M} = \frac{1}{M^2_M - p^2 - I \sqrt{p^2} \Gamma_M}, 
\end{eqnarray}
where masses and widths of mesons are taken from PGD \cite{ParticleDataGroup:2022pth}.   

For the decay $\tau \to K^*_0(1430) \pi \nu_\tau$ the amplitude has a similar structure with the replacement of the vertices $W K^*_0(700) \pi \to W K^*_0(1430) \pi$ , $K_1K^*_0(700) \pi \to K'_1 K^*_0(1430) \pi$ and meson masses $M_{K^*_0(700)} \to M_{K^*_0(1430 )}$. Numerical estimates of the branching fractions are given in Table \ref{tab_2}.
	\begin{figure}[h]
		\center{\includegraphics[scale = 0.8]{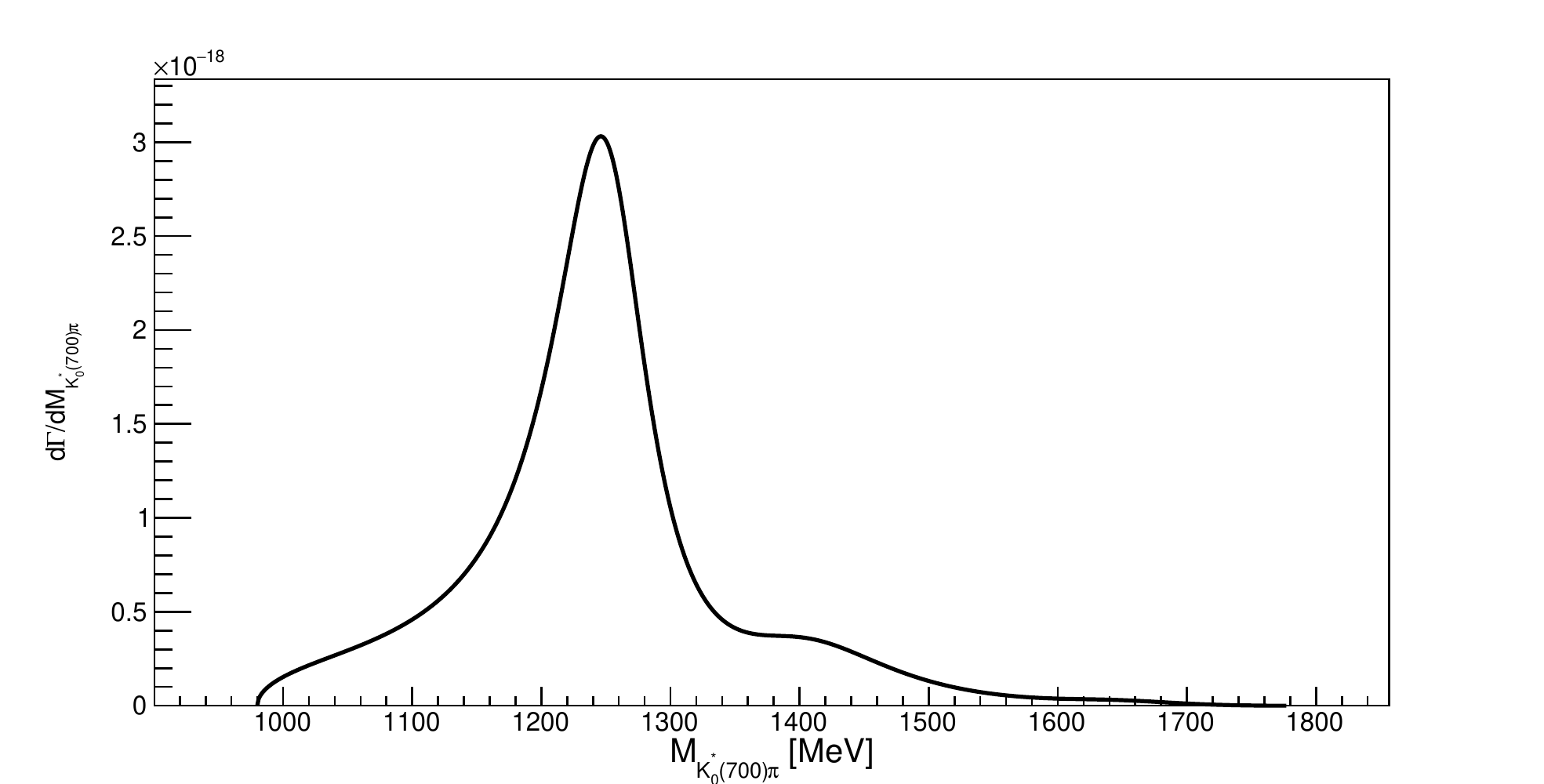}}
		\caption{Invariant mass distribution for the $\tau^- \to K^*_0(700) \pi \nu_\tau$ decay}
		\label{K0pi}
	\end{figure}
	
		\begin{figure}[h]
		\center{\includegraphics[scale = 0.8]{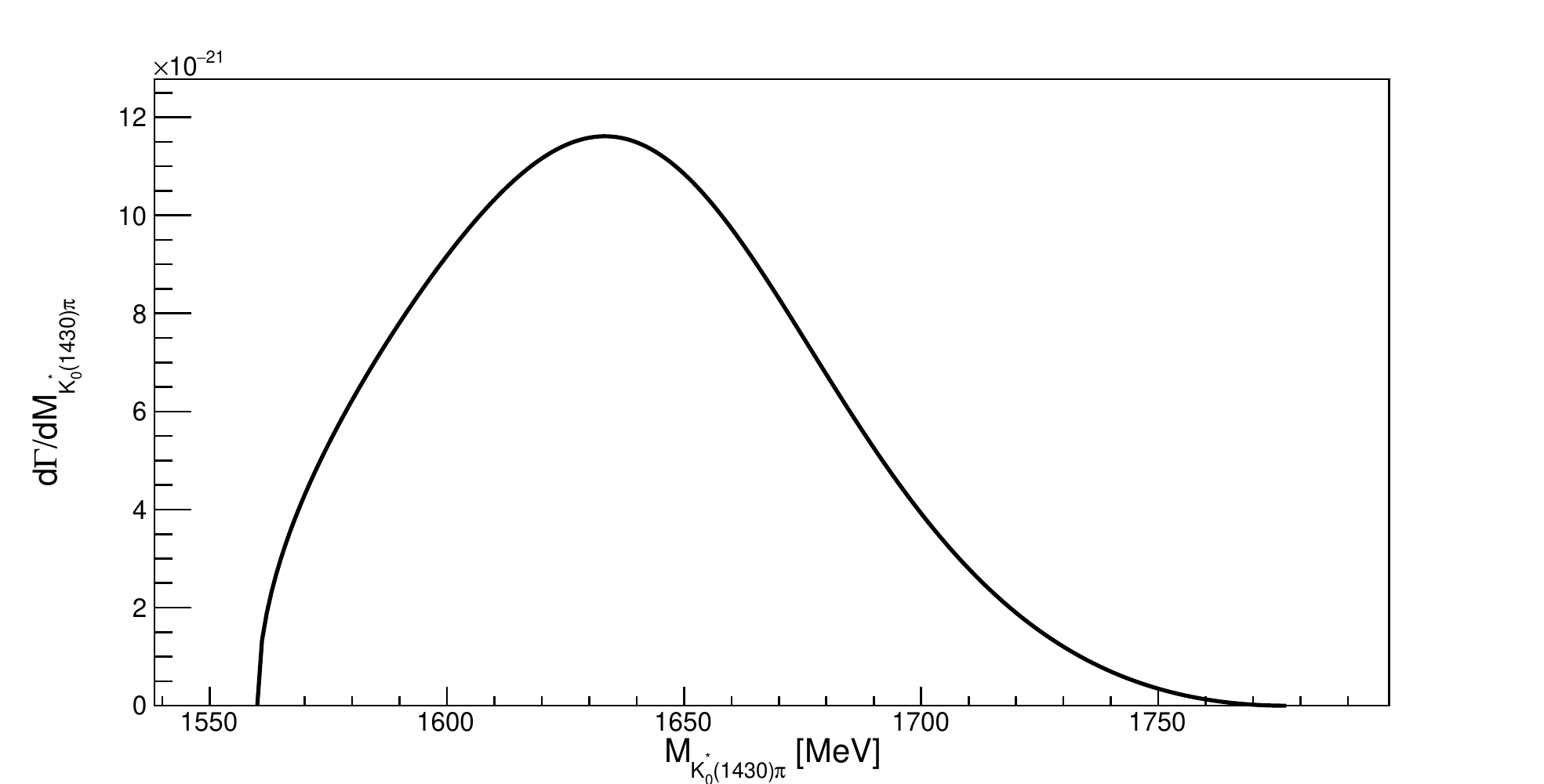}}
		\caption{Invariant mass distribution for the $\tau^- \to K^*_0(1430) \pi \nu_\tau$ decay}
		\label{K0ppi}
	\end{figure}
	
		\begin{figure}[h]
		\center{\includegraphics[scale = 0.8]{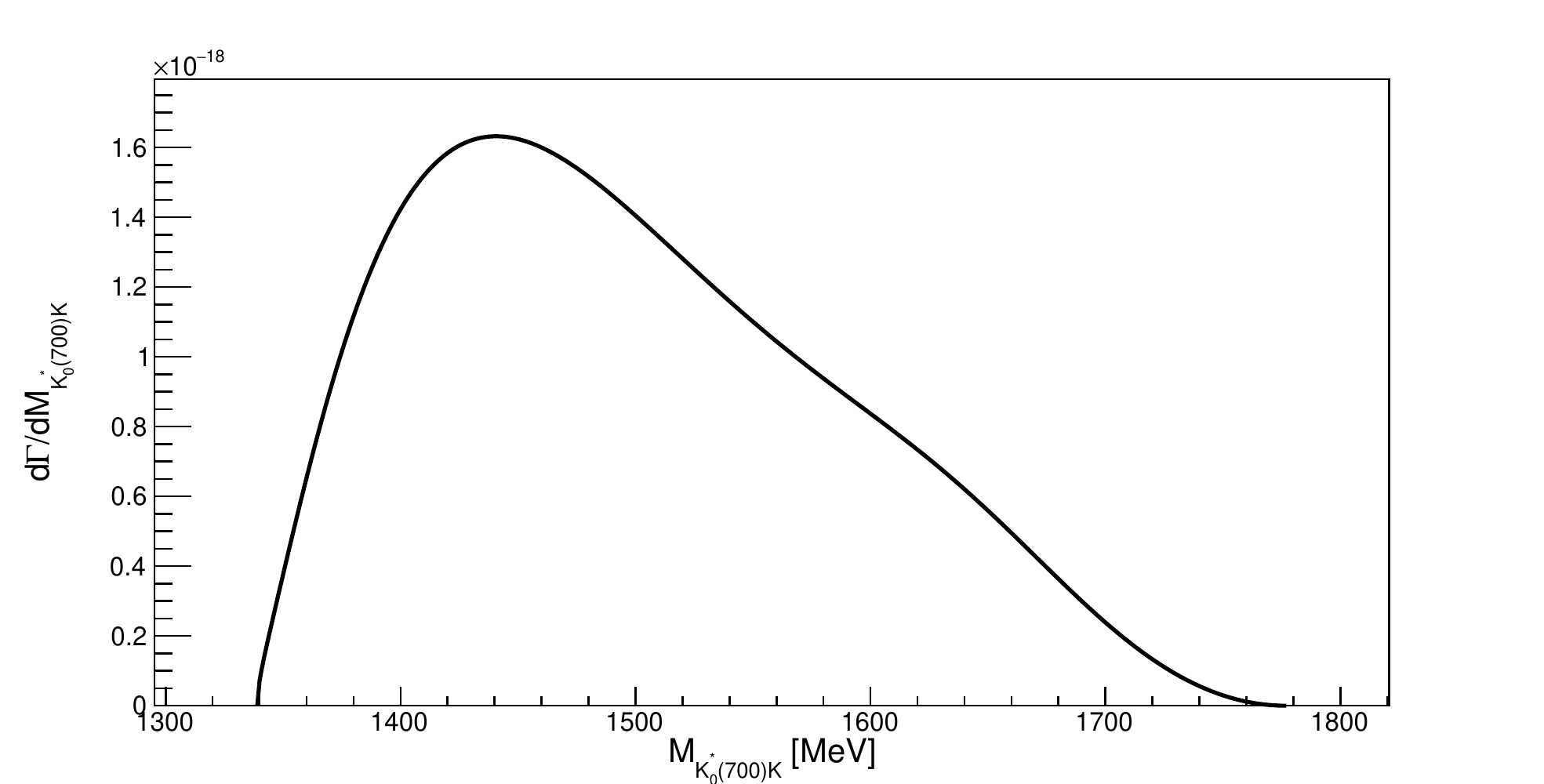}}
		\caption{Invariant mass distribution for the $\tau^- \to K^*_0(700) K \nu_\tau$ decay.}
		\label{K0K}
	\end{figure}
	
Next consider the decay of $\tau \to K^*_0(800) K \pi \nu_\tau$. It should be noted that this decay differs from the decays described above in that it proceeds through intermediate channels involving non-strange mesons $a_1$, $a'_1$, $\pi$ and $\pi'$. As a result, for the total amplitude of this decay, taking into account all contributions in the framework of the extended NJL model, we obtain the following amplitude:
	\begin{eqnarray}
	\label{amplitude_5}
         \mathcal{M}(\tau \to K^*_0(700) K \nu_\tau) & = & 2\sqrt{2} G_F V_{ud} L_\mu \left[ 
         \frac{C_{a_1}}{g_{a_1}} \biggl[ g_{\mu\nu} ( p^2 - 6 m^2_u) - p_\mu p_\nu \biggl] BW_{a_1} I^{a_1 K^*_0 K}_{11}         
           \right. \nonumber\\
         && \left. +  \frac{C_{a'_1}}{g_{a_1}} \biggl[ g_{\mu\nu} \left( p^2 - 6 m^2_u \right) - p_\mu p_\nu \biggl] BW_{a'_1} I^{a'_1 K^*_0 K}_{11}  \right. \nonumber\\    
           && \left. + 4 m_s F_\pi \left( I^{\pi K^*_0 K}_{11} BW_\pi + \gamma_{\pi'} I^{\pi' K^*_0 K}_{11} BW_{\pi'} \right) g_{\mu\nu} p_\nu   
        \right] {\left( p_{K^*_0} - p_K \right)}_\nu,
	\end{eqnarray}
where $\gamma_{\pi'} = F_{\pi'}/F_\pi \approx 0.054$ \cite{Volkov:2022jfr} and $p = p_{K^*_0} + p_K$.  

The amplitude of the decay $\tau \to K^*_0 \eta \nu_\tau$ takes the form
	\begin{eqnarray}
	\label{amplitude_6}
         \mathcal{M}(\tau \to K^*_0(700) \eta \nu_\tau) & = & 8 G_F V_{us} F_K L_\mu \left[ 
         \left( m_s I^{K K^*_0 \eta_u}_{11} - \sqrt{2} m_u I^{K K^*_0 \eta_s}_{11} \right) BW_K       
           \right. \nonumber\\
         && \left. + \gamma_{K'} \left( m_s I^{K' K^*_0 \eta_u}_{11} - \sqrt{2} m_u I^{K' K^*_0 \eta_s}_{11} \right) BW_{K'}    
        \right] {\left( p_{K^*_0} + p_{\eta} \right)}_\mu,
	\end{eqnarray}
here we take into account the $u, d$ and $s$ quark parts of the $\eta$ meson. For the transition $K(K') \to K^*_0 \eta$, we use the vertex obtained in \cite{Volkov:2022ukj}. Using this amplitude for the branching fraction we obtain
 	\begin{eqnarray}
        Br(\tau \to K^*_0(700) \eta \nu_\tau)_{NJL} = 3.599 \times 10^{-8}.
	\end{eqnarray}

\begin{table}[h!]
\begin{center}
\begin{tabular}{ccccccc}
\hline
Decay mode & Contact & AV & AV'  & PS & PS' & Total  \\
\hline
$Br(\tau \to K^*_0(700) \pi \nu_\tau) \times 10^{-4}$  & 0.719 & 4.859 & 0.022 & 0.528 & 0.003 & 4.757 \\
$Br(\tau \to K^*_0(1430) \pi \nu_\tau) \times 10^{-6}$	& 0.085 & 0.027 & 2.09 & 0.008 & 0.079 & 1.801 \\
$Br(\tau \to K^*_0(700) K \nu_\tau) \times 10^{-4}$ & 3.830 & 11.826 & 0.186 & 0.874 & $2.514\times 10^{-4} $ & 5.005 \\
\hline
\end{tabular}
\end{center}
\caption{Branching fractions for $\tau$ lepton decays. The contributions from various channels are shown in different column. Columns AV and $\text{AV'}$ correspond to the diagrams with intermediate axial vector mesons in the ground and excited states. Contributions of pseudoscalar channels are shown as PS and $\text{PS'}$. The column Total contains summary result for the contributions of all diagrams.}
\label{tab_2}
\end{table} 

\section{Conclusions}
As noted in the Introduction, for a correct description of a number of scalar mesons it is necessary to take into account the tetraquark components of these states. This primarily refers to the isovector $a_0(980)$ and isoscalar $f_0(980)$ states. At the same time, within the extended $U(3) \times U(3)$ quark chiral NJL model, it is possible to obtain quite satisfactory results under the assumption of the leading $q\bar{q}$ component of the light $f_0(500)$ meson, strange mesons and all the first radially excited states of the scalar nonet \cite{Volkov:1999yi, Volkov:1998ax}. In our recent paper \cite{Volkov:2022ukj}, it was shown that under this assumption, in the framework of our extended NJL model one can obtain quite satisfactory agreement for the main decays of the $K^*_0(700)$ and $K^*_0(1430)$ mesons with experimental data of the BaBar collaboration \cite{BaBar:2021fkz}. Int the present paper, we continue this study and give a number of predictions for $\tau$ decay branching fractions and differential decay widths in the framework of the extended NJL model.

Our calculations show that the main contribution to the $\tau \to K^*_0(700) \pi \nu_\tau$ decay width comes from the axial vector channel with intermediate mesons $K_1(1270)$ and $K_1(1400)$ considered together with the contribution of the contact diagram with an intermediate axial-vector $W$ boson. In the case of the $\tau \to K^*_0(1430) \pi \nu_\tau$ decay, the channel with intermediate excited $K'_1$ mesons determines the width. In both cases, pseudoscalar channels give a small contribution to the decay width. In determining the width for the $\tau \to K^*_0(700) K \nu_\tau$ decay, the main role is played by the intermediate nonstrange $a_1$ meson with the axial-vector part of the contact diagram. Here the pseudoscalar channel also gives a small contribution to the partial width. The predictions for the differential decay widths are presented in Figures \ref{K0pi}, \ref{K0ppi} and \ref{K0K}. It should be noted that all results are obtained with the previously determined parameters of the NJL model without using any additional free parameters \cite{Volkov:2017arr, Volkov:2022jfr}.

Unfortunately, the strange scalar meson $K^*_0(1430)$ is currently insufficiently studied from an experimental point of view in the low-energy region. However, its mass and width are fairly well defined in heavy meson decays \cite{BaBar:2021fkz, Belle:2004drb, LHCb:2014ioa, LHCb:2019xmb}, where scalar mesons play an important role in accounting for their contribution in intermediate states. Unfortunately, at present there are no direct experimental data on the study of the decays we described. We sure that future experiments on $\tau$ lepton decays at the Super $c - \tau$ factories \cite{Charm-TauFactory:2013cnj, Luo:2018njj} and Belle II \cite{Belle-II:2018jsg} and other experiments will make it possible to study tau decays with sufficiently high statistics.

\subsection*{Acknowledgments}
We are grateful to prof. A.B. Arbuzov for his interest in our work and important remarks which 
improved the paper.

\end{document}